# Transmembrane transport of polymer brush-grafted nanoparticles into giant vesicles*

Shuai He（贺帅）, Junxing Pan（潘俊星）[†], and Jinjun Zhang（张进军）[†]

*Department of Physics and Electronic Engineering, Shanxi Normal University, Taiyuan, 030000, China*

**Abstract:** Polymer brush-grafted nanoparticles have significant application value in fields such as gene therapy and targeted drug delivery. A profound understanding of the interaction mechanisms between these particles and cell membranes represents a critical scientific challenge in biophysics. Using the Self-Consistent Field Theory (SCFT), this work systematically explores the transmembrane transport of polymer brush-grafted nanoparticles into giant vesicles. The impacts of critical parameters—polymer brush grafting density, nanoparticle size, and giant vesicle membrane thickness—on transport behavior are comprehensively elucidated. The findings reveal two distinct transmembrane transport mechanisms for polymer brush-grafted nanoparticles, which are governed by membrane thickness and grafting density. At high grafting density, the nanoparticles undergo direct transmembrane translocation; at low grafting density, transport occurs via endocytosis. Thermodynamic analysis identifies entropy as the dominant driving force for this process.

---

[†] Corresponding author. E-mail: panjx@sxnu.edu.cn
[†] Corresponding author. E-mail: zhangjinjun@sxnu.edu.cn

## I. Introduction

Nanotechnology has been widely applied in biomedical fields, including targeted drug formulation[1-2], disease diagnosis[3-5], and the design of drug delivery systems[6-8]. Compared to conventional therapies, nanotechnology-enabled drugs can more effectively target specific cells while minimizing systemic toxicity. The safe and stable targeted delivery of nanomedicines is critical for improving therapeutic efficacy and accelerating clinical translation. The selection of appropriate carriers and a mechanistic understanding of nanoparticle-cell interface interactions are key determinants for successful nanomedical applications. Recent advancements have been reported in fields such as oral squamous cell carcinoma treatment[9-12], cancer drug resistance inhibition[13-18], vitiligo therapy[19], neuropsychiatric disease intervention[20], and cardiovascular disease management[21].

Among various nanocarriers, polymer brush-grafted nanoparticles (PBNPs) have garnered significant attention due to their high chemical stability, strong environmental responsiveness, versatile synthetic methods, tunable structures, and excellent biocompatibility[22-24]. These carriers can prolong drug retention and accumulation in tumor tissues by evading premature elimination[25-26], facilitate cancer cell tracking and killing, and enable pH-, temperature-, or time-dependent smart drug release[24]. Ballauff et al.[27-30] demonstrated that polymer brushes preserve the structural and functional integrity of adsorbed biomolecules.

Comprehending how nanoparticles interact with cell membranes, and their transmembrane transport mechanisms, is fundamental to deciphering biological complexity and regulating cellular processes. Extensive research has been conducted on the transmembrane transport of particles in previous studies[31-36]. However, the intricate architecture of biological membranes and stringent experimental conditions present challenges to mechanistic investigations. Giant vesicles (GVs)—synthetic lipid structures with micrometer-scale diameters and biomembrane-mimetic properties—offer a versatile platform for bridging fundamental research and clinical applications in biomedicine and nanotechnology[37-40]. These structures provide an ideal model system for investigating cellular processes such as endocytosis and exocytosis.

Their surface modification allows for precise drug targeting in delivery systems, with additional potential in gene therapy and genome editing. Extensive studies have utilized GVs to elucidate nanoparticle-membrane interaction mechanisms. Yang et al.[41] employed computer simulations to investigate the translocation processes of nanoparticles with different shapes (e.g., spheres, ellipsoids, rods, discs, and pushpin-like particles) and sizes across a lipid bilayer. Their findings indicate that a nanoparticle's ability to penetrate a lipid bilayer is determined by its contact area with the bilayer and the local curvature at the point of contact. Bahrami et al.[42] investigated the cooperative phagocytosis of nanoparticles by membranes and vesicles, demonstrating that distant nanoparticles are indirectly engulfed by vesicles driven by curvature effects. Mathieu et al.[43] explored interactions between GVs and charged core-shell magnetic nanoparticles, revealing that these interactions do not stem from a simple electrostatic phenomenon. Smith et al.[44] utilized dissipative particle dynamics (DPD) to analyze nanoparticle-membrane interactions, allowing particles to detach from larger membranes during wrapping. Ou et al.[45] systematically investigated the interactions between sub 10 nm cationic LNPs (cLNPs; e.g., 4 nm in size) and various model cell membranes using molecular dynamics simulations. Their findings demonstrate that the membrane binding of cLNPs is an entropy-driven process, enabling them to differentiate between membranes with different lipid compositions. Ahmad et al.[46] performed thermodynamic analyses of the mutual effects between nanoparticles and membranes. Wang et al.[47] found that the protein corona structure modulates the interfacial properties of nanocarriers, leading to a shift in the translocation mode of liposomes across cell membranes—from an energy-independent membrane fusion mechanism to an energy-dependent endocytosis mechanism. Livadaru et al.[48] employed statistical mechanics to investigate the transmembrane dynamics of peptide-conjugated nanoparticles. Yan[49] and Guo[50] utilized dissipative particle dynamics (DPD) and molecular dynamics (MD), respectively, to investigate the interactions between dendritic polymers and phospholipid bilayers. Yan's findings revealed that initiator density, together with polymerization rates, significantly affects the structure of polymer brushes, initiation efficiency, and grafting density at high initiator densities. Guo demonstrated that three distinct states exist in the interaction dynamics between a dendrimer and a lipid bilayer membrane. Katsov[51] and Zhang[52] further leveraged self-consistent field theory (SCFT) to investigate phospholipid bilayer fusion and nanoparticle-encapsulated liposome coalescence. Additionally, Wang[53] and Yang[54] et al. provided multidisciplinary insights into the transmembrane transport mechanisms of lipid nanoparticles. In summary, the interactions of nanoparticles with cell membranes and their transmembrane transport mechanisms are fundamental to advancing our understanding of biological complexity and cellular processes. Through innovative platforms such as GVs combined with cutting-edge theoretical models—including dissipative particle dynamics (DPD) and molecular dynamics (MD)—researchers have made significant strides in elucidating these intricate interactions, paving the way for breakthroughs in targeted drug delivery systems and gene therapy. Among the various methods for investigating such problems, the mean-field method stands out due to its advantages in analyzing molecular interactions, flexibility in theoretical modeling, and capacity to integrate with experimental data. Xin et al.[55] implemented the charge-self-consistent DFT+DMFT formalism by interfacing a full-potential all-electron DFT code with three hybridization-expansion-based continuous-time quantum Monte Carlo impurity solvers, further demonstrating the strengths of mean-field methods.

In this work, we employ Self-Consistent Field Theory (SCFT) to systematically investigate the interfacial interactions and dynamic self-assembly mechanisms between PBNPs and GVs. By precisely tuning the interparticle distance between PBNPs and GVs, we mapped the spatiotemporal evolution of interactions at the lipid bilayer-nanoparticle interface and correlate these interactions with structural remodeling in the hybrid system. To unravel the thermodynamic driving forces, we constructed a free energy landscape that explicitly decomposes entropic and enthalpic contributions during dynamic assembly. This approach provides mechanistic insights into the equilibrium and non-equilibrium states of the system.

## Ⅱ. Model and Method

The model system (Figure 1) consists of a homopolymer-grafted rigid nanoparticle and a GV, both dispersed in a hydrophilic homopolymer solvent (S). The GV comprises a lipid bilayer formed by two distinct lipid types, A and B, each containing a hydrophilic head group and a hydrophobic tail. In this paper, $d$ represents the distance between the nanoparticle's center and the center of the GV bilayer membrane. The nanoparticle radius is denoted as $R_p$. When the centers of the nanoparticle and the GV bilayer coincide ($d=0$), this point is taken as the origin of coordinates, indicates that the nanoparticle is located at the GV membrane. If $d<0$, this signifies that the nanoparticle is inside the GV.

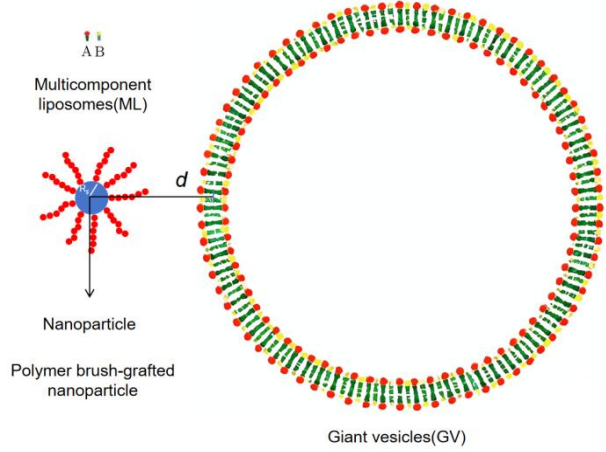

**Figure 1. Schematic illustration of the interaction between a polymer brush-grafted rigid nanoparticle and a GV. Here, $R_p$ denotes the radius of the nanoparticle, and $d$ represents the distance between the center of the nanoparticle and the center of the bilayer membrane.**

In this paper, all the species are incompressible, and their local volume fractions $\phi_i(\mathbf{r})$ satisfy the following relationship,

$$\phi_{hA}(\mathbf{r}) + \phi_{tA}(\mathbf{r}) + \phi_{hB}(\mathbf{r}) + \phi_{tB}(\mathbf{r}) + \phi_{br}(\mathbf{r}) + \phi_{S}(\mathbf{r}) = \phi_{0}(\mathbf{r}) \qquad （1）$$

Here, the value of $\phi_i(\mathbf{r})$ depends on the relative position of $\mathbf{r}$ and the nanoparticle radius $R_p$,

$$\phi_0(\mathbf{r}) = \begin{cases} 0 & |\mathbf{r}| \leq R_p \\ \frac{1}{2}\{1-\cos[\pi(\mathbf{r}-R_p)/\varepsilon]\} & R_p < |\mathbf{r}| < R_p + \varepsilon \\ 1 & |\mathbf{r}| \leq R_p + \varepsilon \end{cases} \quad (2)$$

Where $\varepsilon$ is a sufficiently small value.

The free energy $F$ can be obtained as follows,

$$\begin{aligned}\frac{NF}{\rho_0 k_B TV} =& -\phi_A f_{tA}\ln\left(\frac{\Omega_A}{V\phi_A f_{tA}}\right) - \phi_B f_{tB}\ln\left(\frac{\Omega_B}{V\phi_B f_{tB}}\right) - \phi_S\ln\left(\frac{\Omega_S}{V\phi_S}\right) - \phi_{br}f_{br}\ln\left(\frac{\Omega_{br}}{V\phi_{br}f_{br}}\right) \\ &+\frac{1}{V}\int d\mathbf{r}\begin{bmatrix} \chi_{hA-tA}N\phi_{hA}(\mathbf{r})\phi_{tA}(\mathbf{r}) + \chi_{hA-hB}N\phi_{hA}(\mathbf{r})\phi_{hB}(\mathbf{r}) + \chi_{hA-tB}N\phi_{hA}(\mathbf{r})\phi_{tB}(\mathbf{r}) \\ + \chi_{hA-S}N\phi_{hA}(\mathbf{r})\phi_S(\mathbf{r}) + \chi_{tA-hB}N\phi_{tA}(\mathbf{r})\phi_{hB}(\mathbf{r}) + \chi_{tA-tB}N\phi_{tA}(\mathbf{r})\phi_{tB}(\mathbf{r}) \\ + \chi_{hA-br}N\phi_{hA}(\mathbf{r})\phi_{br}(\mathbf{r}) + \chi_{hB-br}N\phi_{hB}(\mathbf{r})\phi_{br}(\mathbf{r}) + \chi_{hA-S}N\phi_{hA}(\mathbf{r})\phi_S(\mathbf{r}) \\ + \chi_{hB-S}N\phi_{hB}(\mathbf{r})\phi_S(\mathbf{r}) + \chi_{tA-br}N\phi_{tA}(\mathbf{r})\phi_{br}(\mathbf{r}) + \chi_{tB-br}N\phi_{tB}(\mathbf{r})\phi_{br}(\mathbf{r}) \\ + \chi_{S-br}N\phi_S(\mathbf{r})\phi_{br}(\mathbf{r}) - \omega_{hA}(\mathbf{r})\rho_{hA}(\mathbf{r}) - \omega_{tA}(\mathbf{r})\rho_{tA}(\mathbf{r}) - \omega_{hB}(\mathbf{r})\rho_{hB}(\mathbf{r}) \\ - \omega_{tB}(\mathbf{r})\rho_{tB}(\mathbf{r}) - \omega_{br}(\mathbf{r})\rho_{br}(\mathbf{r}) - \omega_S(\mathbf{r})\rho_S(\mathbf{r}) \\ - H(\mathbf{r})N(\phi_{hA}(\mathbf{r})+\phi_{hB}(\mathbf{r})-\phi_{tA}(\mathbf{r})-\phi_{tB}(\mathbf{r})+\phi_S(\mathbf{r})+\phi_{br}(\mathbf{r})) \\ -\xi(\mathbf{r})(\phi_0(\mathbf{r})-\phi_{hA}(\mathbf{r})-\phi_{tA}(\mathbf{r})-\phi_{hB}(\mathbf{r})-\phi_{tB}(\mathbf{r})-\phi_S(\mathbf{r})-\phi_{br}(\mathbf{r})) \end{bmatrix}\end{aligned} \quad (3)$$

where $k_B$ is the Boltzmann constant, $T$ is the temperature, $N$ is the polymeric degree, $\rho_0^{-1}$ represents the volumes of lipid tail segment and solvent chain segment, and $V$ is the total volume of the system, $f_{ti}(i=A/B)$ is volume fraction of lipid $i$ tail, $f_{br}$ is volume fraction of polymer brush. The first four terms on the right hand side of eq.(3) represent the entropy contributions of lipids, polymer brush and solvents. $\Omega_i (i = A \text{ or } B)$, $\Omega_{br}$ and $\Omega_S$ are the single-molecule partition functions of lipid $i$, polymer brush and solvent S:

$$\Omega_i = \int d\mathbf{r} q_i(\mathbf{r},s)q_i^+(\mathbf{r},s) \quad (4)$$

$$\Omega_{br} = \int d\mathbf{r} q_{br}(\mathbf{r},s)q_{br}(\mathbf{r},1-s) \quad (5)$$

$$\Omega_S = \int d\mathbf{r} q_S(\mathbf{r},s)q_S(\mathbf{r},1-s) \quad (6)$$

where $s$ is the segment index of the lipid molecules, from the end of one tail $s = 0$ to the end of the other tail $s = 1$, the lipid head is in the middle of two tails at position $s = 1/2$. The end-segment distribution functions $q_i(\mathbf{r},s)$, $q_i^+(\mathbf{r},s)$, $q_s(\mathbf{r},s)$, and $q_{br}(\mathbf{r},s)$ obey the following modified diffusion equations:

$$\frac{\partial q_i}{\partial s} = \frac{Na_0^2}{6}\nabla^2 q_i - \omega_{ti} q_i \tag{7}$$

$$\frac{\partial q_i^+}{\partial s} = \frac{Na_0^2}{6}\nabla^2 q_i + \omega_{ti} q_i \tag{8}$$

$$\frac{\partial q_S}{\partial s} = \frac{Na_0^2}{6}\nabla^2 q_S - \omega_S q_S \tag{9}$$

$$\frac{\partial q_{br}}{\partial s} = \frac{Na_0^2}{6}\nabla^2 q_{br} - \omega_{br} q_{br} \tag{10}$$

The 5-17 terms in eq.(3) represent the contribution of interaction energy between different components. The Flory-Huggins interaction energy of different components are expressed by $\chi_{ij}$. The 18-23 terms represent the contribution of the mean fields acting on different components, where $\omega_{hi}$, $\omega_{ti}$, $\omega_s$, and $\omega_{br}$ represent the fields acting on the head groups($hi$), tail segments($ti$), polymer brush segments($br$) and solvent segments($S$) respectively, $\rho_i$ ($i=h_A$, $h_B$, $t_A$, $t_B$, $br$, $S$) represents the number density of different components. $H(\mathbf{r})$ term represents the field of nanoparticle acting on the lipids, polymer brush and solvents:

$$H(\mathbf{r}) = \begin{cases} \infty & |\mathbf{r}| \leq R_p \\ \frac{\Lambda}{\varepsilon}\{1 + \cos[\pi(\mathbf{r}-R_p)/\varepsilon]\} & R_p < |\mathbf{r}| < R_p + \varepsilon \\ 0 & |\mathbf{r}| \leq R_p + \varepsilon \end{cases} \tag{11}$$

where $\Lambda$ is the interaction strength between the particle surface and the components in the system, $\varepsilon$ is a sufficiently small value. $\xi(\vec{r})$ in the last term of eq.(3) is a Lagrange multiplier to ensure the incompressibility of the system.

The initial condition for the above four diffusion equations(7)-(10) are $q_i(\mathbf{r},0)=1$, $q_i^+(\mathbf{r},0)=\exp(-\omega_{hi}(\mathbf{r}))$, $q_s(\mathbf{r},0)=1$, $q_{br}(\mathbf{r},0)=1$. Minimizing the free energy in eq.(3) with respect to the mean fields and monomer densities, we can obtain a set of mean-field equations:

$$\omega_{hi}(\mathbf{r}) = \gamma_i \chi_{hi-hj} N\phi_{hj}(\mathbf{r}) + \chi_{hi-ti} N\phi_{ti}(\mathbf{r}) + \chi_{hi-tj} N\phi_{tj}(\mathbf{r}) \\ + \chi_{hi-S} N\phi_S(\mathbf{r}) + \chi_{hi-br} N\phi_{br}(\mathbf{r}) - H_p(\mathbf{r})N + \xi(\mathbf{r})  \quad (12)$$

$$\omega_{ti}(\mathbf{r}) = \chi_{ti-hi} N\phi_{hi}(\mathbf{r}) + \chi_{ti-hj} N\phi_{hj}(\mathbf{r}) + \chi_{ti-tj} N\phi_{tj}(\mathbf{r}) + \chi_{ti-S} N\phi_S(\mathbf{r}) \\ + \chi_{ti-br} N\phi_{br}(\mathbf{r}) - H_p(\mathbf{r})N + \xi(\mathbf{r}) \quad (13)$$

$$\omega_S(\mathbf{r}) = \chi_{hi-S} N\phi_{hi}(\mathbf{r}) + \chi_{ti-S} N\phi_{ti}(\mathbf{r}) + \chi_{hj-S} N\phi_{hj}(\mathbf{r}) + \chi_{tj-S} N\phi_{tj}(\mathbf{r}) \\ + \chi_{br-S} N\phi_{br}(\mathbf{r}) - H_p(\mathbf{r})N + \xi(\mathbf{r}) \quad (14)$$

$$\omega_{br}(\mathbf{r}) = \gamma_i \chi_{hi-br} N\phi_{hi}(\mathbf{r}) + \chi_{ti-br} N\phi_{ti}(\mathbf{r}) + \gamma_i \chi_{hj-br} N\phi_{hj}(\mathbf{r}) + \chi_{tj-br} N\phi_{tj}(\mathbf{r}) \\ + \chi_{br-S} N\phi_S(\mathbf{r}) - H_p(\mathbf{r})N + \xi(\mathbf{r}) \quad (15)$$

$$\rho_{hi}(\mathbf{r}) = \frac{\phi_i f_{ti} V}{\Omega_i} q_i(\mathbf{r},1) q_i^+(\mathbf{r},1) \quad (16)$$

$$\phi_{hi}(\mathbf{r}) = \gamma_i \rho_{hi}(\mathbf{r}) \quad (17)$$

$$\phi_{ti}(\mathbf{r}) = \frac{\phi_i f_{ti} V}{\Omega_i} \int_0^1 ds\, q_i(\mathbf{r},1) q_i^+(\mathbf{r},1) \quad (18)$$

$$\phi_S(\mathbf{r}) = \frac{\phi_S V}{\Omega_S} \int_0^1 ds\, q_S(\mathbf{r},1) q_S^+(\mathbf{r},1-S) \quad (19)$$

$$\phi_{br}(\mathbf{r}) = \frac{\phi_{br} f_{br} V}{\Omega_i} \int_0^1 ds\, q_{br}(\mathbf{r},1) q_{br}^+(\mathbf{r},1) \quad (20)$$

where $\gamma_i (= V_{hi}/(N\rho^{-1}))$ is the ratio of the head volume to the tail volume.

We assume $\chi_{hA-tB}N=\chi_{hB-tA}N=\chi_{s-tA}N=\chi_{s-tB}N=\chi_{hA-br}N=\chi_{hB-br}N=\chi_{s-br}N=\chi_{tA-br}N=\chi_{tB-br}N$ =20, $\chi_{hA-s}N=\chi_{hB-s}N=0$, $\chi_{hA-tA}N=\chi_{hB-tB}N=3$ for convenient calculations. All of the sizes and volumes involved in the system are respectively measured in units of $a_0$ and $\rho_0^{-1}$, and we set $a_0=1$, $\rho_0^{-1}=1$. The chain contour length for the lipid tails, polymer brush and solvents are discretized into 50 segments. The simulation computations are based on periodic boundary conditions. The above self-consistent equations are solved using real space methods [56,57].

## III. RESULTS AND DISCUSSION

Firstly, we focus on investigating the regulatory roles of two core factors — polymer brush grafting density and phospholipid molecule concentration — in the transmembrane transport mechanisms of nanoparticles. As shown in Figure 2, the transmembrane behavior of PBNPs exhibits two distinct mechanisms: direct membrane transport and endocytic membrane transport. In Figure 2a, the polymer brush concentration and phospholipid grafting density corresponding to direct membrane transport and endocytic membrane transport are provided. The cyan region corresponds to direct transmembrane transport, while the red region corresponds to endocytic transmembrane transport. This figure shows the effects of phospholipid concentration and grafted polymer brush density on the membrane transport mechanism. Low phospholipid molecule concentration（$\phi_A = \phi_B < 0.07$）, nanoparticles predominantly undergo direct transmembrane transport into the vesicle (Figure 2c). At high phospholipid molecule concentration ($\phi_A = \phi_B > 0.07$), endocytic membrane transport occurs when the grafting density of polymer brushes is low (Figure 2e); direct membrane transport occurs when the grafting density is high. Direct membrane transport involves the opening of a channel on the GVs, allowing PBNPs to enter directly. Endocytic membrane transport involves the separation of some phospholipid molecules on the GVs and the opening of a channel to enable PBNPs to enter. The transmembrane mechanism of PBNPs is determined by the grafting density of polymer brush and the volume fraction of phospholipid molecules. Lower grafting density of polymer brush and higher volume fraction of phospholipid molecules favor endocytic membrane transport. An increase in the grafting density of polymer brushes and a decrease in the volume fraction of phospholipid molecules lead to a transition from endocytic membrane transport to direct transmembrane transport. This transition is mainly attributed to the increased particle size caused by the elevated grafting density. Larger particles induce membrane wrapping, which stretches the membrane, reduces the system entropy, and raises the energy barrier for this process. Thus, this pathway becomes unfavorable. A decrease in membrane thickness is consistent with this mechanism.

It is worth noting that the grafting density of polymer brushes affects the stability of the transport process. When the grafting density is below 0.004, rupture of the grafted polymer brushes is observed in both transport mechanisms (Figures 2b, d). The rupture of the polymer brush membrane is influenced by the grafting density of polymer brushes and the interaction between PBNPs and vesicles, as shown in Figure 2f. When the grafting density of polymer brushes is 0.004, increasing the attraction strength leads to the occurrence of membrane rupture. When the interaction is fixed, the phenomenon of membrane rupture disappears as the grafting density of polymer brushes increases.

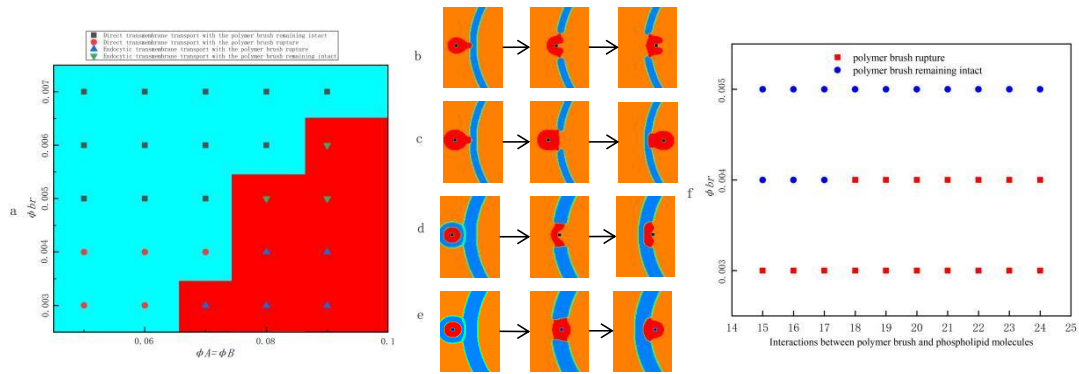

**Figure 2. (a) Phase diagram: The cyan region represents direct transmembrane transport, and the red region represents endocytic transmembrane transport. (b) Direct transmembrane transport with the polymer brush rupture. (c) Direct transmembrane transport with the polymer brush remaining intact. (d) Endocytic transmembrane transport with the polymer brush rupture. (e) Endocytic transmembrane transport with the polymer brush remaining intact. (f) Phase diagram: Interactions between polymer brushes and phospholipid molecules result in the rupture of polymer brush membranes.**

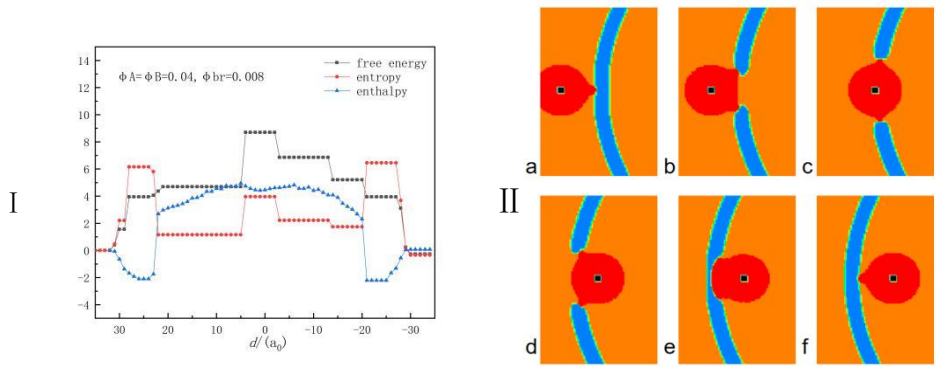

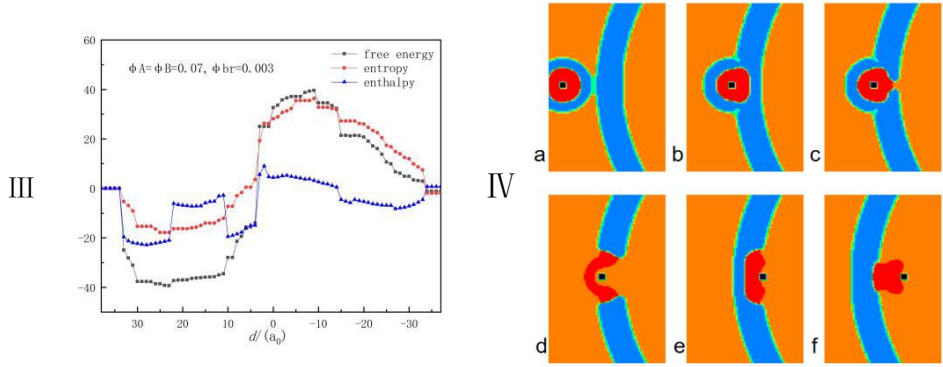

**Figure 3. (I) Curves showing changes in free energy, entropy, and enthalpy during the direct transmembrane transport process. (II) Local morphological map illustrating the interaction between nanoparticles and GVs during direct transmembrane transport. (III) Curves of free energy, entropy, and enthalpy changes during the endocytic transmembrane process. (IV) Local morphological map depicting the interaction between nanoparticles and giant vesicles during endocytosis.**

To systematically investigate the two transmembrane transport mechanisms, we calculated the changes in the system's free energy, entropy, and enthalpy during both transport processes. The results are shown in Figure 3. Figure I illustrates the variation curves of free energy, entropy, and enthalpy during direct transmembrane transport. Figure Ⅱ presents a partial schematic diagram corresponding to the interaction between nanoparticles and GVs as depicted in Figure I. Here, the concentration of phospholipid molecules is 0.04, and the grafting density of polymer brushes is 0.008. When nanoparticles grafted with polymer brush approach the membrane, the system's configurational entropy increases. This is because the polymer brush disrupts the conformational order of the GVs. As the $d$ decreases, such disruption inevitably enhances the configurational entropy. When $d$ reaches a certain position, the channel in the GVs opens, leading to strong interactions between the polymer brush and lipid molecules. This restricts their motion, leading to a strictly confined conformational state and a significant reduction in configurational entropy. When the nanoparticles penetrate to the midpoint of the membrane, the membrane's compression effect on the polymer brushes increases slightly, causing a small rise in configurational entropy. Once the nanoparticles cross the midpoint, the compression effect weakens, resulting in two minor decreases in configurational entropy. When the vesicle channel closes, the conformational relationship between the nanoparticles and vesicles becomes similar to that at the initial stage of transmembrane transport. At this point, most of the nanoparticles are located inside the GVs, leading to a significant increase in configurational entropy. As the transmembrane process progresses, the nanoparticles fully enter the GVs, the configurational entropy decreases and gradually stabilizes. On the other hand, In the initial stage of membrane penetration ($d = 32$—23), PBNPs interact with GVs. Both the polymer brushes and the phospholipid heads of the vesicles are hydrophilic, leading to mutual attraction and a consequent decrease in enthalpy. During membrane penetration ($d = 22$—7), a channel forms in the vesicle membrane. At this stage, interactions between the polymer brushes and the hydrophobic phospholipid tails dominate, inducing repulsion between the two components and a corresponding increase in enthalpy. When $d = 6$—0, the channel in the GVs reaches its maximum opening: the direction of the repulsive force exerted by the phospholipid tails reverses, the repulsion weakens while the attraction persists, resulting in a slight decrease in enthalpy. As $d$ ranges from 0 to -6, the vesicle channel closes gradually, which enhances the repulsive force of the phospholipid tails on the polymer brushes and leading to a slight rise in the system's enthalpy. For $d = -7$ to $-19$, the progressive closure of the vesicle channel weakens the interaction between the phospholipid tails and polymer brushes, causing a stepwise decrease in enthalpy. In the final stage of membrane penetration ($d = -20$ to $-29$), the vesicle channel fully closes, terminating the interaction between the phospholipid tails and polymer brushes and eliminating the repulsive force. This triggers an immediate and significant drop in the system's enthalpy. Subsequently, the PBNPs gradually enter the GVs, and the contact between the polymer brushes and phospholipid heads decreases. The system's enthalpy then increases gradually until the nanoparticles fully enter the vesicle and attain thermodynamic stability. By analyzing the changes in entropy and enthalpy, we can derive the free energy changes of the system. Throughout the entire transport process, entropy exerts a greater impact on free energy than enthalpy. These findings are consistent with the conclusion in Yang work[55] that nanoparticle transport processes are entropy-driven.

Figures 3III and 3IV show changes in free energy, entropy, and enthalpy during endocytic transport, as well as the corresponding schematics diagrams, respectively. Here, the concentration of phospholipid molecules is 0.07, and the grafting density of polymer brushes is 0.003. When nanoparticles grafted with polymer brushes approach GVs, the system's configurational entropy decreases. This is driven by two factors: first, the disruption of the conformational order between polymer brushes and GVs; second, the strong interaction between polymer brushes and lipid head groups, which pull some lipid molecules away from the vesicle membrane. These lipids molecules effectively encapsulate the polymer brushes and further reduce the configurational entropy by restricting the motion of brushes. With the decrease in $d$, more lipid molecules are incorporated into the GVs, leading to an increase in configurational entropy until the channel closes and the entropy reaches its maximum value. As transmembrane transport proceeds, the interaction between the PBNPs and GVs weakens gradually, resulting in a decrease in configurational entropy. This decline persists until the nanoparticles fully enter the GVs, the configurational entropy stabilizes. On the other hand, Where $d$ ranges from 34 to 22, nanoparticles are wrapped by phospholipids molecules, forming a hydrophilic layer composed of phospholipid heads. This layer interacts with the hydrophilic outer membrane of the vesicle, leading to a decrease in enthalpy. When $d$ ranges from 21 to 11, the wrapped nanoparticles merge into the vesicle membrane. At this point, the hydrophilic polymer brushes interact with the hydrophobic phospholipid tails, causing an increase in enthalpy. As the channel opens ($d$ from 10 to 4), the enthalpy drops sharply, followed by a slight rise. This is due to the combined effect of the interaction between the polymer brushes and phospholipids, as well as membrane deformation. At $d = 3$, the phospholipids wrapping the nanoparticles dissociate suddenly, causing a sharp increase in enthalpy. From $d = 2$ to -10, the channel remains open, allowing the polymer brushes to interact with the phospholipid tails and resulting in a slight decrease in enthalpy. When the channel is closed ($d$ from -10 to -27), the interaction between the polymer brushes and the hydrophobic phospholipid tails persists, leading to a further decrease in enthalpy. As the nanoparticles are about to enter the vesicle membrane (*d* from -27 to -33), these polymer brushes interact with the hydrophilic heads of the inner membrane; this interaction causes the enthalpy to increase. Once the transport process is complete, the enthalpy remains stable. Regarding the changes in entropy, the wrapping of nanoparticles by phospholipids molecules reduces the conformational entropy. As the transport process proceeds, the system's conformational entropy increases continually. The opening of the channel, dissociation of phospholipids, and closing of the channel collectively contribute to an increase in entropy. This process continues until the nanoparticles fully enter the vesicle, after which the entropy stabilizes gradually. By analyzing the entropy and enthalpy at different stages, we find that the system's free energy follows a trend of "decrease-increase-decrease-stabilization".

## IV. CONCLUSIONS

This study adopts the self-consistent field method to model the interactions between PBNPs and GVs. Two distinct transmembrane transport mechanisms are observed, namely direct transmembrane transport and endocytic transmembrane transport. Direct membrane transport refers to the opening of a channel on the GVs, which allows PBNPs to enter directly and complete transmembrane transport. Endocytic membrane transport involves the separation of some phospholipid molecules on the GVs to first encapsulate PBNPs, followed by the opening of a channel on the GVs to complete transmembrane transport. At low phospholipid molecule concentration ($\phi_A = \phi_B < 0.07$), only direct membrane transport occurs; at high phospholipid molecule concentration ($\phi_A = \phi_B > 0.07$) and low grafting density of grafted polymer brushes, endocytic membrane transport occurs, and as the grafting density increases, the membrane transport mode changes to direct membrane transport. It is worth noting that when the grafting density ranges from 0.003 to 0.004, the interaction between polymer brushes and phospholipid molecules is strong, and the polymer brush membrane will rupture during the membrane transport process; when the interaction is weak, the polymer brush membrane with a grafting density of 0.004 will not rupture during the membrane transport process. When the grafting density of polymer brushes is high ($\phi_{br} > 0.005$), the polymer brush membrane does not rupture during the membrane transport process under both mechanisms. Through the analysis of the changes in free energy, entropy, and enthalpy curves, we can gain a deeper understanding of the occurrence of the two membrane transport mechanisms, which both originate from the synergistic effect of entropy and enthalpy. These findings provide a reference for the design of nanodrug delivery materials and facilitate an in-depth understanding of the interactions between cell membranes and substances.